\documentclass[%
 reprint,
 superscriptaddress,
 amsmath,amssymb,
 aps,
 prc,
floatfix,
]{revtex4-1}

\usepackage{graphicx}
\usepackage{dcolumn}
\usepackage{ulem}
\usepackage{bm}
\usepackage{lipsum}
\usepackage{subcaption}
\usepackage{caption}
\usepackage{xcolor}
\usepackage{float}

\def\cj#1#2{\sout{#1}{\textcolor{red}{#2}}}
\usepackage{multirow}

\begin{document}


\title{Data based two-body current contribution to neutrino-nucleus cross section}

\author{Tomasz Bonus}
\email{tomasz.bonus@ift.uni.wroc.pl}
 \affiliation{Institute of Theoretical Physics, University of Wroc{\l}aw, Plac Maxa Borna 9, 50-204, Wroc{\l}aw, Poland}
\author{Jan T. Sobczyk}
 \email{jan.sobczyk@uwr.edu.pl}
 \affiliation{Institute of Theoretical Physics, University of Wroc{\l}aw, Plac Maxa Borna 9, 50-204, Wroc{\l}aw, Poland}
\author{Micha\l\ Siemaszko}
 \affiliation{Institute of Theoretical Physics, University of Wroc{\l}aw, Plac Maxa Borna 9, 50-204, Wroc{\l}aw, Poland}
 \author{Cezary Juszczak}
 \affiliation{Institute of Theoretical Physics, University of Wroc{\l}aw, Plac Maxa Borna 9, 50-204, Wroc{\l}aw, Poland}
\date{\today}

\begin{abstract}
A phenomenological model of two-body current (2p2h) contribution to neutrino cross section is introduced. Predictions of the Valencia model for 2p2h \cite{Nieves:2011pp} are modified using recent CC$0\pi$ measurements from T2K and MINERvA experiments. Our results suggest a significant increase of the 2p2h cross section at neutrino energies bigger than 1~GeV and also a redistribution of 2p2h events as function of energy and momentum transfer. This may have a big impact on neutrino energy reconstruction  in neutrino oscillation parameters.
\end{abstract}
\maketitle


\section{Introduction}
\label{sec:introduction}

One of the most important unknowns in modeling 
neutrino-nucleus cross sections~\cite{Alvarez-Ruso:2017oui} is the size of the contribution coming from two-body current (2p2h) mechanism~\cite{Marteau:1999jp, Martini:2009uj, Martini:2010ex}. 
It is important to have a precise estimate of the fraction of events  originating from this mechanism because in detectors  like SuperKamiokande they cannot be distinguished from charge current \mbox{quasi-elastic} (CCQE) scatterings on bound nucleons
\begin{equation}
\label{eq:ccqe}
\nu_l \ +\ n\ \rightarrow\ l^- \ +\ p, \ \ \ \ \ \bar{\nu_l} \ +\ p\ \rightarrow\ l^+ \ +\ n,
\end{equation}
where $l$ is lepton's flavor, n, p are neutron and proton, respectively. This leads to a bias in the neutrino energy reconstruction~\cite{Sobczyk:2012ms, Martini:2012fa, Lalakulich:2012ac, Mosel:2013fxa, Nikolakopoulos:2018sbo, Ankowski:2014yfa, Ankowski:2016bji} and strongly affects the precision of neutrino oscillation parameters measurements.

Over the last decade a lot of  theoretical studies were done aiming to understand the situation~\cite{Nieves:2011pp, Megias:2016fjk, VanCuyck:2016fab, VanCuyck:2017wfn, Rocco:2015cil, Rocco:2016ejr, Lovato:2017cux, Rocco:2018mwt, Rocco:2018tes, Mosel:2017ssx}. The most reliable ab initio computations exist only on a restricted phase space and for light nuclei. At larger neutrino energies theoretical model predictions differ significantly among themselves~\cite{Dolan:2019bxf}.

Experimental studies focus mainly on CC$0\pi$ (called also \mbox{CCQE-like}) measurements with
the signal defined as `no pion in the final state` \cite{AguilarArevalo:2010zc}.
Most of the CC$0\pi$ events originate from the CCQE mechanism, but there is  a significant contribution from the two-body current mechanism and also from pion production with consequent absorption inside nucleus.
The advantage of this type of measurements comes from simplicity of the definition of experimental signal. The data analysis does not depend on uncertain predictions for the hadrons in the final state. The available theoretical models for the 2p2h contribution give predictions for the final state lepton only and modeling final state hadrons is based on  approximations~\cite{Sobczyk:2012ms} and nucleon final state interactions effects \cite{Niewczas:2019fro}. 

Recent CC$0\pi$ measurements were done by T2K and MINERvA experiments. In both cases results are published in a form of flux averaged double differential cross section in kinematic variables describing final state \hbox{(anti-)muon}. T2K measured cross section  for neutrinos and antineutrinos on hydrocarbon \cite{Abe:2016tmq, Abe:2020jbf} and on water \cite{Abe:2017rfw, Abe:2019sah}.  \hbox{MINERvA} published measurements on hydrocarbon for antineutrinos~\cite{Patrick:2018gvi} and neutrinos~\cite{Ruterbories:2018gub}. 
Altogether, there is a lot of information in the experimental data that has not yet been fully explored. The most important CCQE contribution to the CC$0\pi$ signal is well understood thanks to electron scattering studies. It has been established that in the typical T2K kinematical region theoretical models used in neutrino community reproduce the QE peak region quite well~\cite{Sobczyk:2017mts}. For the pion production and absorption there have been many studies which put a lot of constraint on them~\cite{Gonzalez-Jimenez:2017fea, Sobczyk:2018ghy}. The most uncertain is the 2p2h contribution and a natural question arises how much can be learnt about it from the CC$0\pi$ measurements.

The goal of this paper is to answer this question and as a result to propose a new phenomenological mode of 2p2h. 
The computations are done using a NuWro Monte Carlo event generator~\cite{Golan:2012wx}, but our procedure  is quite general and can be employed in other MC generators and be used in neutrino oscillation experimental studies. 

Our study is inspired by the MINERvA experiment attempt to resolve events' kinematics completely with calorimetric-type measurement of the interacting (anti-)neutrino energy~\cite{Rodrigues:2015hik, Gran:2018fxa}.
A study done in the context of GENIE Monte Carlo (MC) generator~\cite{Andreopoulos:2009rq} allowed to identify a kinematical region where more strength from the 2p2h mechanism is needed, relative to predictions of the Valencia theoretical model~\cite{Nieves:2011pp, Gran:2013kda}. Contrary to the above mentioned study our work uses information contained in the final state muon only. 

Our paper is organized as follows.
In section~\ref{sec:approach}  our approach  is presented and  the data sets used in the numerical analysis are described. 
Section~\ref{sec:results} outlines our main results: the new model and its performance compared to the experimental data. 
The Section~\ref{sec:conclusions} contains a discussion of the results and final remarks. 
Appendices A and B include technical details supplementary to the Section~\ref{sec:approach}. A simple toy model illustrates our method of analyzing the data based on a separation of the covariance matrices into shape and normalization parts.

\section{Our approach}
\label{sec:approach}

The starting point for our investigation is the Valencia  model of the 2p2h contribution described in Ref.~\cite{Nieves:2011pp} with a restriction on the values of momentum transfer \hbox{$q\leq 1.2$~GeV/c~\cite{Gran:2013kda}}. It is implemented in NuWro in terms of five structure functions depending on  energy and momentum transfers ($\omega,q$). The Valencia model does not provide predictions for final state nucleons and this information is added in a factorization scheme using a model proposed in Ref.~\cite{Sobczyk:2012ms}. The structure functions $W_j$ define double differential cross section in final state lepton kinematical variables:

\begin{equation}
\frac{d^2\sigma^{mec}}{d\omega dq}  
= \frac{G_F^2\cos^2\theta_c q}{2\pi
E^2}L_{\mu\nu}W^{\mu\nu}
\end{equation}
where 
\begin{eqnarray}
L_{\mu\nu}W^{\mu\nu}
=&W_1(Q^2+m^2)\nonumber\\ 
+ &W_2\left( 2E(E-\omega )-\displaystyle{\frac{m^2+Q^2}{2}}\right)\nonumber\\
\pm &\displaystyle{\frac{W_3}{M}} \left( EQ^2-\displaystyle{\frac{\omega}{2}}(m^2+Q^2) \right)\nonumber\\
+&\displaystyle{\frac{W_4}{M^2}} \left( {\frac{1}{2}}Q^2m^2+{\frac{1}{2}}m^4\right)-\displaystyle{\frac{W_5}{M}}m^2E
\label{eq:inclusive}
\end{eqnarray}

In the above equations $E$ is neutrino energy, $m$ is charged lepton mass, $G_F$ is Fermi constant, $\theta_c$ - Cabibbo angle,  $Q^2\equiv q^2-\omega^2$ and $M$ is nucleon mass.  A sign $\pm$ in the $W_3$ containing term refers to neutrino/antineutrino cases. At neutrino energies in current and planned short- and long-baseline oscillation experiments the contributions from $W_4$ and $W_5$ containing terms are strongly suppressed due to presence of charged lepton mass in a multiplicative factor.

Our considerations are based on the hypothesis that the 
overall double differential cross section defined by the Valencia model should be scaled by an unknown function $S(\omega, q)$:
\begin{equation}
\frac{d^2\sigma^{mec, phenom}}{d\omega dq}  = \frac{G_F^2\cos^2\theta_c q}{2\pi E^2}L_{\mu\nu}W^{\mu\nu} S(\omega, q)
\label{eq:rescaling}
\end{equation}
and a form of $S(\omega , q)$ will be deduced from the CC$0\pi$ data. Equivalently, this may be viewed as a simultaneous rescaling of all the structure functions $W_j$  by $S(\omega, q)$:

\begin{equation}
    W_j(\omega, q)\rightarrow \tilde W_j(\omega, q)= W_j(\omega, q) S(\omega, q)
\end{equation}

Even if the assumption introduced in Eq.~\ref{eq:rescaling} looks general it is in fact quite restrictive. The proposed rescaling is independent on neutrino energy and is the same for both neutrinos and antineutrinos. In Sect.~\ref{sec:conclusions} we will explain how it can be made more general and realistic.
The form of the scaling function $S(\omega, q)$ will be determined by minimization of $\chi^2$ estimator introduced in \ref{sec:estimator}. 

\subsection{Data sets}

We investigate information from the T2K and MINERvA CC$0\pi$  measurements in the balanced way. Both are done on the same target but with different beams peaked at $\sim 600$~MeV for T2K and $\sim 3.5$~GeV for MINERvA. In the case of MINERvA we include  results from neutrinos \cite{Ruterbories:2018gub} and antineutrinos \cite{Patrick:2018gvi}.   In the case of T2K we include neutrino and antineutrino measurements from Ref.~\cite{Abe:2020jbf}.  In all the considered measurements the results for double differential cross section is reported together with the covariance matrix $V_{j,k}$. In the case of T2K data we use two separate covariance matrices for neutrino and antineutrino results in the same way in which the MINERvA data is available. We disregard the T2K neutrino/antineutrino covariance matrix in order to treat both experiments in a symmetric way.

MINERvA $\nu_{\mu}$ data contains 156 2D bins. They are distributed on 2-dimensional grid of the size 12 by 13. The binning is done respectively by longitudinal (range from 1.5 to 15 GeV/c) and transversal (range from 0 to 2.5 GeV/c) components of the outgoing muon momentum. For the MINERvA $\bar{\nu_{\mu}}$ data the division is done by using the same kinematic variables but binning is different (for the transversal component the range is from 0 to 1.5 GeV/c) resulting in a 10 by 6 grid i.e. 60 2D bins. In the MINERvA experiment there is a limited acceptance of muons: its angle must be lower than $20^o$ with respect to neutrino beam.

T2K data represent double differential  cross section in (anti-)muon momentum and cosine of the lepton scattering angle. The binning is the same for neutrino and antineutrino. Altogether, there are 58 2D bins in each case. Muon momentum range is from 0 to 5 GeV/c. The full range of the cosine is employed. However, in the forward muon directions the binning is much finer. All the backward directions are contained in just one cosine bin extending from -1 to 0.2.

\subsection{NuWro}

NuWro \cite{nuwro} is a neutrino Monte Carlo generator developed at the Wroc\l aw University starting from 2005. It can be used for neutrino energy range from \hbox{$\sim 100$~MeV} to \hbox{$\sim 100$~GeV}. For neutrino-nucleon scattering NuWro uses three interaction modes: CCQE \cite{LlewellynSmith:1971zm} (and elastic for neutral current reactions), RES \cite{Sobczyk:2004va, Graczyk:2009qm} which covers a region of invariant hadronic mass $W\leq 1.6$~GeV and DIS including shallow and deep inelastic processes with $W>1.6$~GeV. In the case of neutrino-nucleus scattering two new interaction modes are: coherent pion production (COH) and two-body current (2p2h).

Simulations done in this paper were done using NuWro version 19.02. Nucleus is treated as a local Fermi gas (LFG). 2p2h events were generated with Valencia model \cite{Nieves:2011pp, Gran:2013kda}. Final state interactions play an important role for RES events and are modelled with Oset et al model \cite{Salcedo:1987md, Golan:2012wx}.

NuWro predictions $\sigma^{model}_k$ in each bin k is a sum of three contributions

\begin{eqnarray}
\sigma^{model}_k= \sigma^{ccqe}_k+\sigma^{res+dis}_k+\sigma^{2p2h}_k
\end{eqnarray}


\subsection{Estimator}
\label{sec:estimator}

Schematically, our estimator is defined as:

\begin{eqnarray}
\chi^2=\sum_{I=1}^4 \chi^2_{I, cov}
\end{eqnarray}
where
\begin{eqnarray}
\chi^2_{I, cov}=\sum_{k,l}(\sigma^{data}_k-\sigma^{model}_k) V^{-1}_{I; k,l} (\sigma^{data}_l-\sigma^{model}_l).
\label{eq:schem}
\end{eqnarray}
$k,l$ run over bins in double differential cross sections and $V_{I;k,l}$ is a covariance matrix for the experiment $I$, \hbox{$I=1,...,4$}.

It turns out that the function $S(\omega, q)$ obtained by minimizing Eq.~\ref{eq:schem} leads to a drastic and clearly nonphysical reduction of the cross section far below the measured cross section in most of the bins. We recognized this behavior as a manifestation of Peelle`s Pertinent Puzzle (PPP) \cite{PPP}. We checked that this effect comes from both MINERvA data sets. Various remedies were proposed to deal with this problem. We decided to follow the ideas proposed in \cite{D'Agostini:1993uj, Aguilar-Arevalo:2013dva}. The overall covariance matrix is decomposed into 'shape', 'normalization' and 'mixed' parts \cite{Aguilar-Arevalo:2013dva}, see the  details in Appendix A. Our estimator for the \hbox{MINERvA} data is constructed as a sum of contributions from shape and  normalization uncertainties. This is similar to the treatment discussed in Ref. \cite{Aguilar-Arevalo:2013dva}. However, while in the MiniBooNE paper only the diagonal part of the shape covariance matrix is explored we include the complete information contained there. We performed several tests of the performance of this method and results are summarized in Appendix B. The final form of our estimator is:

\begin{eqnarray}
\chi^2_{final} = (\chi^2_{shape} + {\cal N})_{\text{MINERvA} \text{ }\nu_\mu} \nonumber \\
+ (\chi^2_{shape} + {\cal N})_{\text{MINERvA} \text{ } \bar{\nu}_\mu}\\
+ (\chi^2_{cov})_{\text{T2K} \text{ }\nu_\mu} + (\chi^2_{cov})_{\text{T2K} \text{ }\bar{\nu}_\mu} \nonumber
\label{eq:chifinal}
\end{eqnarray}
where

\begin{eqnarray}
&\chi^2_{shape}=\nonumber \\
&\displaystyle{\sum_{k,l}}(\sigma^{data}_k-\sigma^{model}_{norm, k}) V_{shape, k,l}^{pseudoinv} (\sigma^{data}_l-\sigma^{model}_{norm, l}).
\label{eq:shape}
\end{eqnarray}
$\sigma^{model}_{norm, l}$ are linearly rescaled model predictions satisfying 

\begin{eqnarray}
\sum_j\sigma_j^{data}=\sum_j\sigma^{model}_{norm, j}.
\end{eqnarray}
$V_{shape, k,l}^{pseudoinv}$ is Moore-Penrose pseudoinverse matrix \cite{Penrose:1955vy} to the `shape` component of the covariance matrix.
${\cal N}$ is defined as

\begin{eqnarray}
{\cal N}=\frac{\displaystyle \left( \sum_k\sigma^{data}_k - \sum_l\sigma^{model}_l\right)^2}{\displaystyle \delta\sigma^2_{norm}}
\end{eqnarray}
with
\begin{eqnarray}
\delta\sigma^2_{norm}\equiv\sum_{j,k}V_{j,k}.
\end{eqnarray}
For the details about the estimator defined in Eq.~\ref{eq:shape} see Appendices A and B. 

$\chi^2_{final}$ is a function of $S(\omega,q)$ and we are looking for its minimum. In the numerical computations we approximate $S(\omega,q)$ by a 2D step function i.e. by a discrete set of values $S_{mn}$ where $m,n$ refer to bins in the $(\omega, q)$ plane. $m,n$ run values $1, \ldots, 24$. Continuity constraints are imposed on values of  $S_{m,n}$ which as a result cannot be changed in a completely random  way, see Sec.~\ref{sec:continuity}.



\subsection{Fitter}

A minimum of $\chi^2_{final}$ was found using a fitter based on a concept of genetic evolution algorithm \cite{Holland:1992:ANA:531075}. 
It was chosen 
because of its flexibility and ability to escape from local minima.


At the beginning all the matrix entries describing parameters $S_{mn}$ are equal one, meaning no scaling whatsoever.
At every iteration the fitter produces a set of 500 matrices, called generation. In each generation the matrices are 
sorted according to the values of $\chi^2_{final}(S_{mn}^k)$. $10\%$ of best performing matrices (the smallest $\chi^2_{final}$) are copied to the next generation as they are.




$80\%$ of the next generation is populated with the offspring from the previous one. In order to produce the offspring, two matrices are selected at random with a probability to select a matrix $S^k$ being:

\begin{equation}
p(S^k) = \frac{\chi^2_{max}-\chi^2_k}{\displaystyle\sum_{i=1}^{500} (\chi^2_{max}-\chi^2_i )}
\label{eq:offspring}
\end{equation}
where $\chi^2_i$ is a value of $\chi^2_{final}$ of $i^{th}$ matrix and $\chi^2_{max}$ is the maximal value of $\chi^2_{final}$ in the generation. 

From these two parents a new matrix is built. In the first step its elements are taken from either of the parents with relative probabilities proportional to those given by Eq.~\ref{eq:offspring}. Continuity constrain is not yet checked at this point.
In the second step about $5\%$ (an exact number is sampled from binomial distribution) of the new matrix entries are selected at random to be modified. The modification is done with $50\%$ of probability either by multiplication factor 
or by addition of a number. Multiplication factor is selected from a normal distribution centered at 1.0 and with a standard deviation 1.0. Negative values are excluded. In the case of addition a number is selected from a normal distribution centered at 0 with standard deviation 0.5. After every single modification is applied it is checked if the new number satisfies the continuity constraint defined in Eq.~\ref{eq:continuity}. If the constraint is not satisfied, the value is changed to the biggest/lowest allowed one.


The last $10\%$ of the new generation consists of randomly generated matrices. They are created in the following way. We start with two empty matrices, $A$ and $B$. Entries of the matrix $A$ are filled with random values selected from a uniform distribution with minimum/ maximum being the lowest/highest values out of all entries from all the previous generations. Once matrix $A$ is constructed, its entries are checked for the continuity constraint. If a given value\cj{s}{} satisfies the constraint, it is copied to the matrix $B$. If the constraint is not satisfied, a minimal/maximal allowed value according to whether the constraint is broken from above or from below, is inserted to matrix $B$ instead. The order in which the values are checked is irrelevant as they are checked within original matrix $A$, which remains unchanged, and the constraint is symmetrical. At the end of this procedure the matrix $B$ is added to the built generation. 

All the percentage values, population sizes etc. were optimized during trial and error process of testing the performance of the algorithm. A lower bound value of 0.1 was imposed as a lowest possible bin value to prevent vanishing cross section from any region. Too high probability of bin modification led to instability and very slow convergence. Lower percentage of matrices copied to next generation slowed process of escaping from local minima. Higher number of random matrices does not help much, as we only need an access to explore new promising regions and the process of investigating them is time consuming. 

\subsection{Continuity constraints}
\label{sec:continuity}

To prevent obtaining rescaling matrices with large differences between neighbouring bins values we added the following constraint allowing for a control of the smoothness of the final matrix:

\begin{equation}
\alpha_{min}\cdot \max_{<k,l>}(S_{kl}) \leq  S_{ij} \leq  \alpha_{max}\cdot \min_{<k,l>}(S_{kl})
\label{eq:continuity}
\end{equation}
where:
\begin{itemize}
    \setlength\itemsep{0em}
    \item $\alpha$ is a user given parameter with a value from the range [0;1),
    \item $\alpha_{min}=1-\alpha$,
    \item $\alpha_{max} = 1 / \alpha_{min}$,
    \item $k,l$ go through the 4 closest neighbours of the $i,j$ bin.
\end{itemize}

\begin{figure}[h!]
    \centering
    \includegraphics[width=0.97\linewidth]{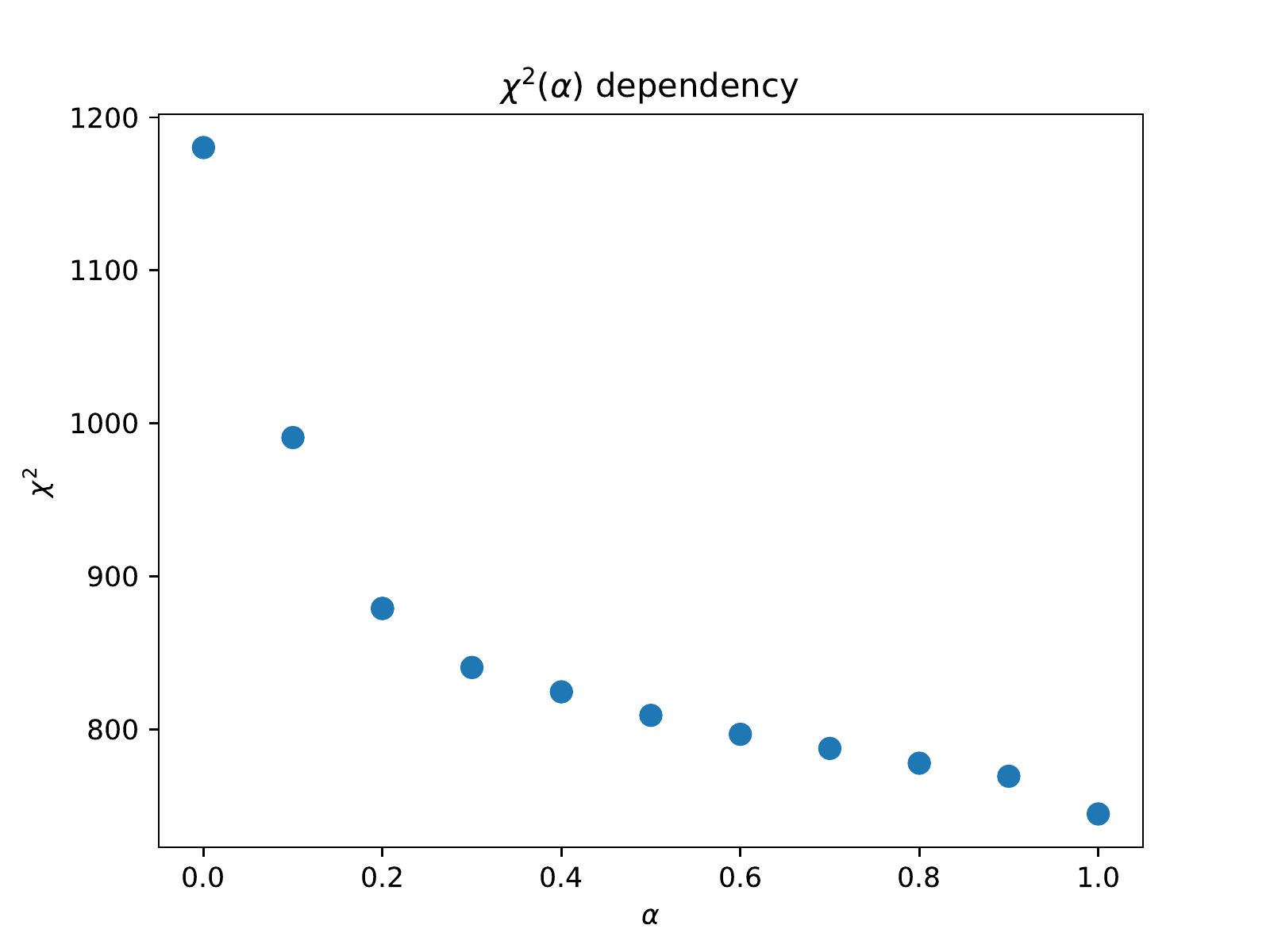}
    \caption{Value of $\chi^2$ as function of $\alpha$ parameter}
    \label{fig:lcurve}
\end{figure}

An impact of changing the values of $\alpha$ on the best fit value of $\chi^2_{final}$ is shown on Fig.~\ref{fig:lcurve}. When we weaken the continuity constraint ($\alpha\rightarrow 1$) the value of $\chi^2_{final}$ at the best fit point becomes smaller. The value $\alpha=0$ corresponds to no rescaling at all. The values $\alpha\neq 1$ 
ensure\cj{s}{} smooth and more physical scaling without sharp and narrow peaks in neighbouring bins. 

A computation for each value of $\alpha$ was performed in $10^5$ iterations. The calculations for 10 values of $\alpha$ on CPU with 6 cores and 12 threads (2 fits were running simultaneously and matrix multiplication was parallelized to achieve 100\% of CPU utilization) takes about 4 hours. The computations were performed 10 times and it was checked that the differences between obtained values of $\chi^2_{final} $ for each $\alpha$ were lower than $1\%$. The best results for each $\alpha$ were chosen as the final result.

The optimal value of $\alpha$ is evaluated by looking at the behavior of  the function defined as

\begin{eqnarray}
(\chi^2(\alpha=0)-\chi^2(\alpha))\cdot (1-\alpha).
\end{eqnarray}
It has a maximum at $\alpha\approx 0.2$ and it is the value used in all further considerations.

\bigskip

\begin{figure}[ht]
    \centering
    \includegraphics[width=0.8\linewidth]{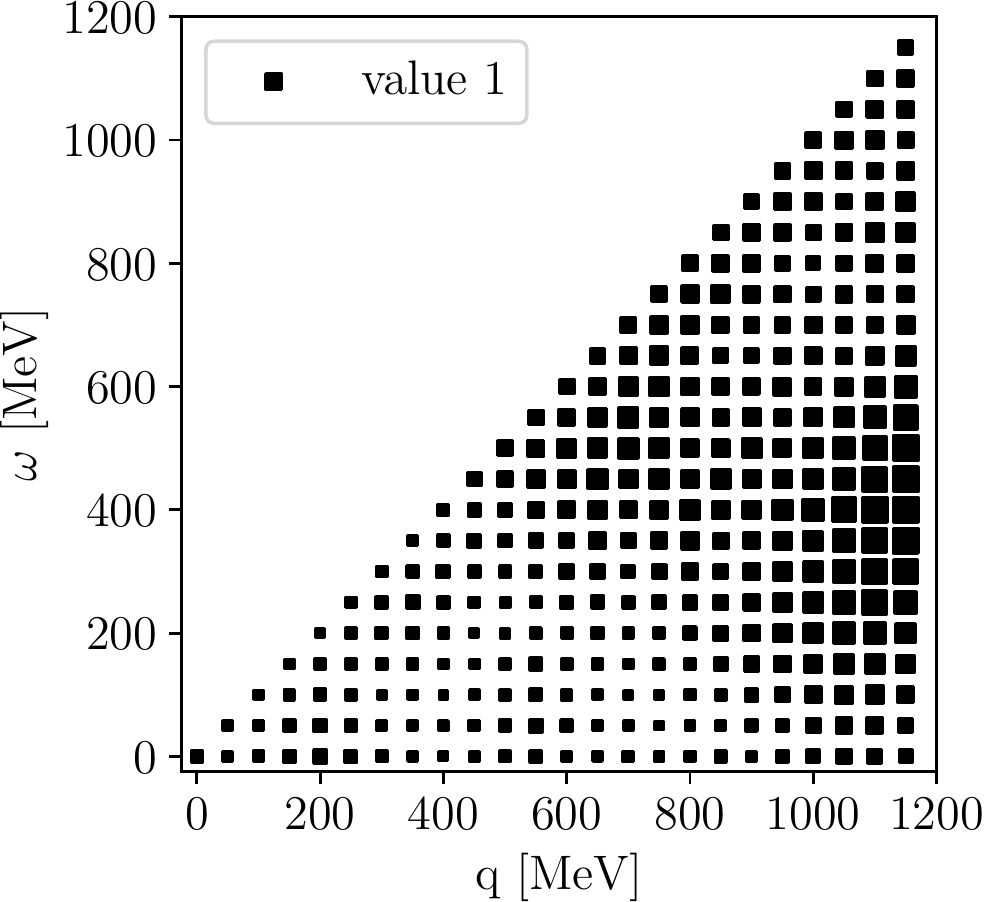}
    \caption{The obtained scaling function for $\alpha = 0.2$ }
    \label{fig:scaling_matrix}
\end{figure}

\section{Results}
\label{sec:results}

\begin{figure*}
    \centering
    \begin{subfigure}{.49\textwidth}
      \centering
      \includegraphics[width=0.9\linewidth]{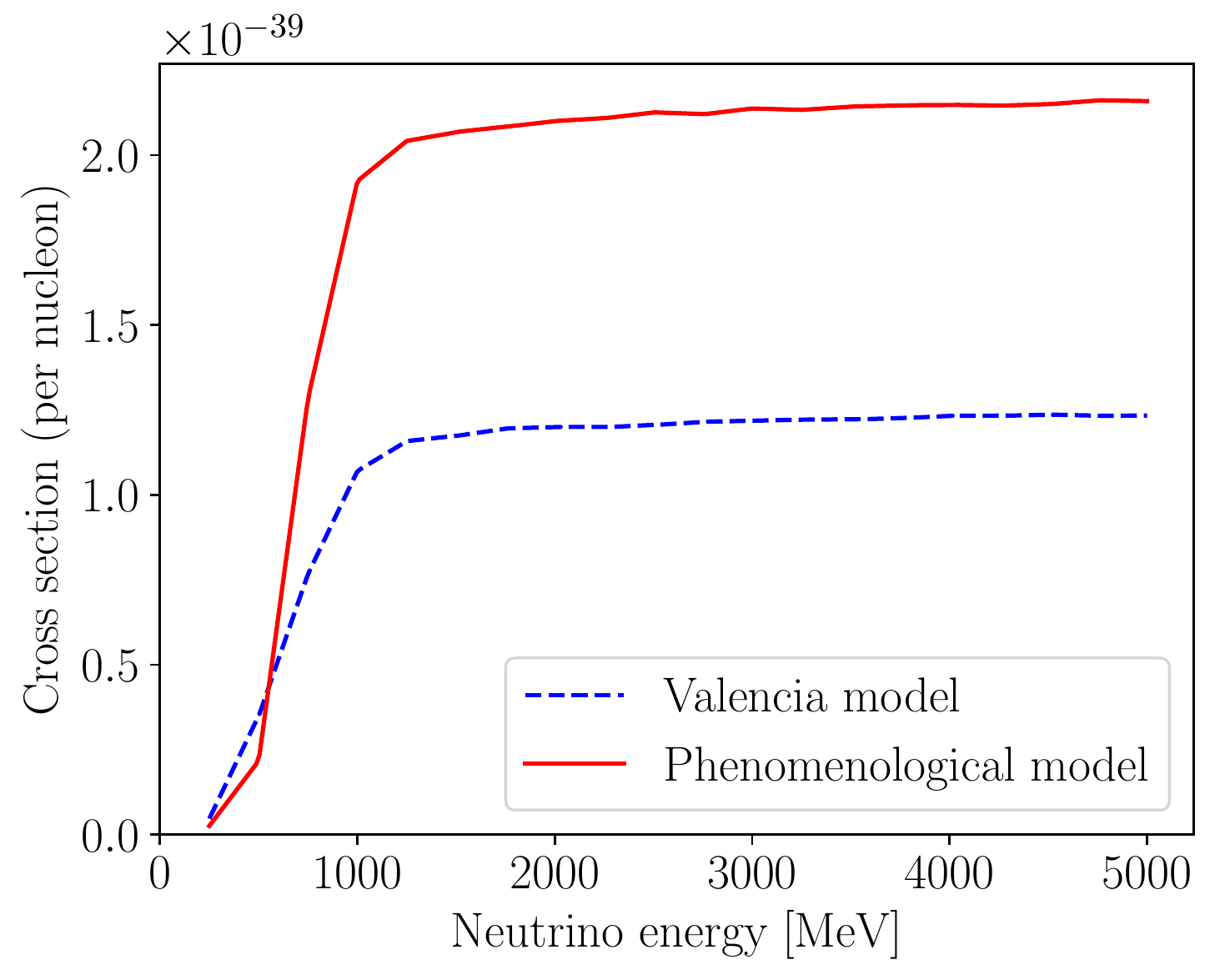}
    \end{subfigure}%
    \medskip
    \begin{subfigure}{.49\textwidth}
      \centering
      \includegraphics[width=0.9\linewidth]{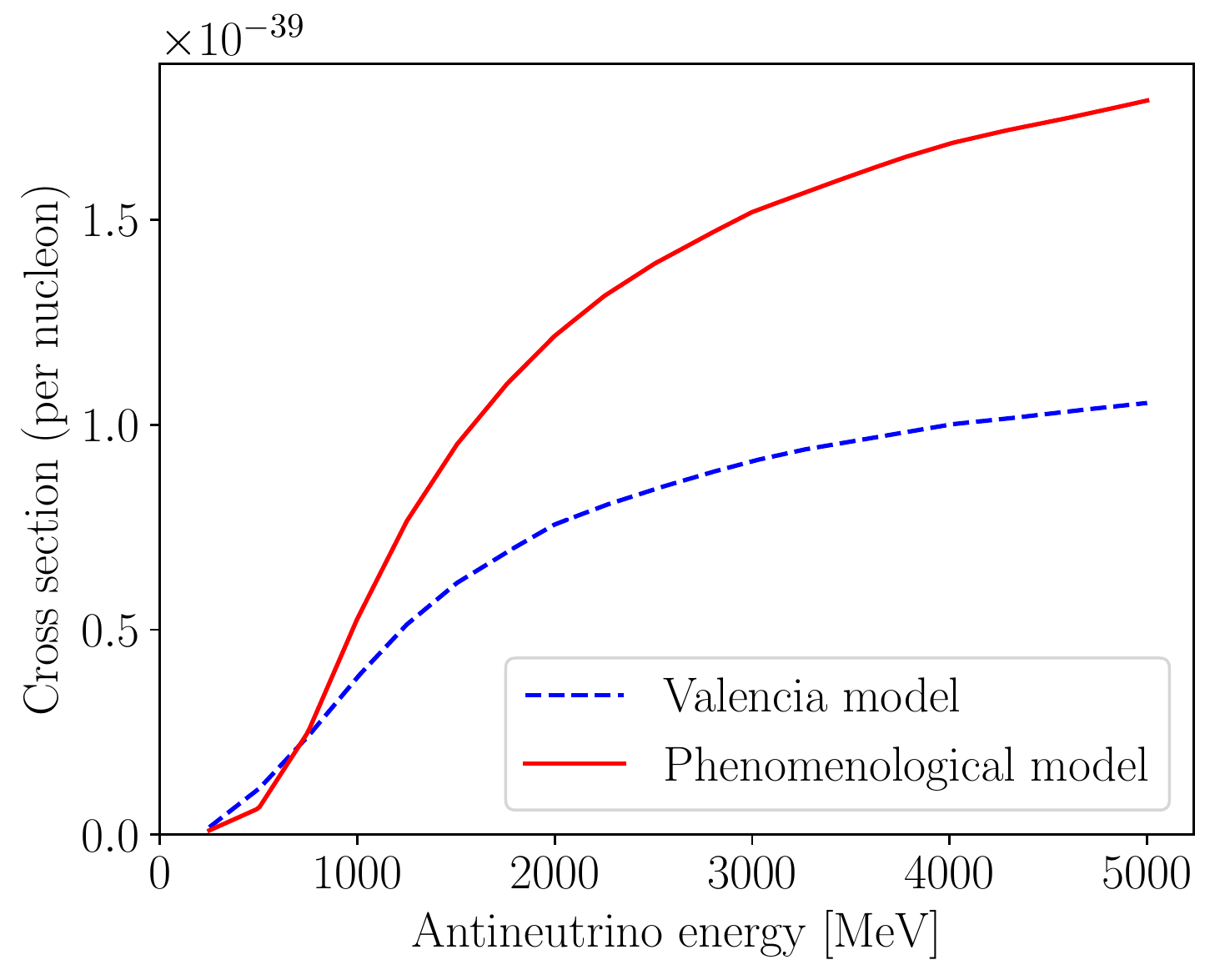}
    \end{subfigure}
    \caption{Cross section dependence on energy.}
    \label{fig:cross_section_energy}
\end{figure*}

The final result for $\alpha = 0.2$ is shown in Fig.~\ref{fig:scaling_matrix}. There are two regions where the scaling makes the 2p2h contribution bigger. The first one is for maximal values of momentum transfer $q\sim 1200$~MeV/c and energy transfer $\omega\in (300,500)$~MeV, and the second one for $q\sim (600,700)$~MeV/c and $\omega \sim (500,600)$~MeV. A reduction of the 2p2h contribution is obtained in a region of lower values of energy transfer. This has important consequences seen in Fig. \ref{fig:cross_section_energy}. There are two (anti-)neutrino energy regimes. For the energies lower than $\sim 700$~MeV the phenomenological model cross sections are lower with respect to the Valencia model. For larger energies the opposite is true and phenomenological model cross sections become bigger. At larger energies the difference becomes vary large and amounts to about 80\%.  As a result the overall T2K cross sections are not changed much while MINERvA cross sections are strongly increased.

\begin{table}[htp]
    \centering
    \begin{tabular}{|c|c|c|c|}
        \hline
        Experiment & D.O.F. & Non-scaled & Scaled  \\
        \hline
        MINERvA $\nu_{\mu}$ & 156 & 618.0+0.8  & 403.0+0.1 \\
        MINERvA $\bar{\nu_{\mu}}$ & 60 & 96.7+1.6 & 132.2+0.2 \\
        T2K $\nu_{\mu}$ & 58 & 262.5 & 137.0  \\
        T2K $\bar{\nu_{\mu}}$ & 58 & 200.8 & 206.6  \\
        \hline
        Sum & 332 & 1180.3 & 879.1 \\
        \hline
    \end{tabular}
    \caption{Contributions to $\chi^2_{final}$ (see Eq. \ref{eq:chifinal}) from each experiment before and after rescaling.}
    \label{tab:chi2}
\end{table}

The contributions from four experiments to the overall value of $\chi^2_{final}$ are listed in Table \ref{tab:chi2}. In the columns 2-4 are shown: numbers of bins in each experiments, values of $\chi^2$ before rescaling and values of $\chi^2$ after rescaling. For the MINERvA experiment we show separately contributions from `shape` and `normalization`, see Eq.~\ref{eq:chifinal}. We see that the final results seem to be determined by the MINERvA neutrino results with the largest number of bins. The contribution to $\chi^2_{final}$ from the MINERvA neutrinos was reduced by a factor of 1/3 at the expense of the MINERvA antineutrino contribution which was increased. This is a signal that our model is not general enough to accomodate both neutrino and antineutrino results at different energies. Still, the overall reduction of the $\chi^2$ is large which means that the new model agrees with the data much better. Another observation is that our method produces an improvement for neutrinos, regardless of their energies but is less successful for antineutrinos. It is a signal that the $W_3$ response function (see Eq.~\ref{eq:inclusive}) which contributes with a different sign for neutrinos and antineutrinos should be rescaled separately.

In Fig.~\ref{fig:wqplane} we show contributions to the cross section from 2p2h events before and after rescaling for each experiment separately. For all the experiments we observe a significant redistribution of the strength always to the region of momentum transfer $\sim 700$~MeV/c and energy transfer $\sim 500$~MeV. For T2K it gives rise to a completely new picture. In the case of T2K neutrinos a region with a large cross section at momentum transfer $\sim 300$~MeV/c and energy transfer $\sim 100$~MeV mostly disappears and similar is the case of neutrinos with smaller values of energy end momentum transfers. The region of the strongest rescaling seen in Fig.~\ref{fig:scaling_matrix} is not a very relevant one for all the experiments and leaves no visible trail in Fig.~\ref{fig:wqplane}.

Another illustration of the performance of our model is seen in Fig.~\ref{fig:bin_cross_sections}. A few typical histograms with experimental results and errors and also model predictions without and with rescaling calculated in this paper are shown together. We see that for MINERvA the overall size of 2p2h contribution is larger than for T2K because of bigger neutrino energy. In the case of T2K the rescaling does not introduce much change. Contrary to that in the case of MINERvA results rescaling makes the overall cross section much bigger. For neutrinos a very good data/MC agreement is obtained while for antineutrinos the rescaling seems to be too strong and MC predictions 
exceed the data points in some bins.

As an additional test we compared the values of $\chi^2$ without correlations. This comparison is closest to the intuitive (sometimes misleading, though) assessment `by eye` of data/MC agreement. In Table~\ref{tab:chi2_std} we see that after rescaling the overall agreement is  much better and the improvement comes mostly from neutrinos. For antineutrinos the model predictions are not changed much but also slightly improved. Apart from MINERvA neutrinos the values of $\chi^2$ after rescaling are close to the number of degrees of freedom 
which means that the agreement is very good.

\begin{table}[]
    \centering
    \begin{tabular}{|c|c|c|c|}
        \hline
        Experiment & D.O.F. & Non-scaled & Scaled  \\
        \hline
        MINERvA $\nu_{\mu}$ & 156 & 462.8  & 358.2 \\
        MINERvA $\bar{\nu_{\mu}}$ & 60 & 65.1 & 62.2 \\
        T2K $\nu_{\mu}$ & 58 & 143.7 & 83.9  \\
        T2K $\bar{\nu_{\mu}}$ & 58 & 101.2 & 98.0  \\
        \hline
        Sum & 332 & 772.8 & 619.6 \\
        \hline
    \end{tabular}
    \caption{Values of $\chi^2$ without the covariance matrix for each experiment before and after rescaling.}
    \label{tab:chi2_std}
\end{table}

\section{Discussion and final remarks} \label{sec:conclusions} 

In this paper we propose a procedure to construct a phenomenological model of two-body current contribution to (anti-)neutrino cross section. A universal rescaling function to be applied to the predictions of the Valencia model~\cite{Nieves:2011pp} is found. Our result is specific to carbon target and also to a selection of models used in numerical computations in NuWro.

\begin{figure*}[t]
    \centering
    \begin{subfigure}{.49\textwidth}
      \centering
      \includegraphics[width=0.8\linewidth]{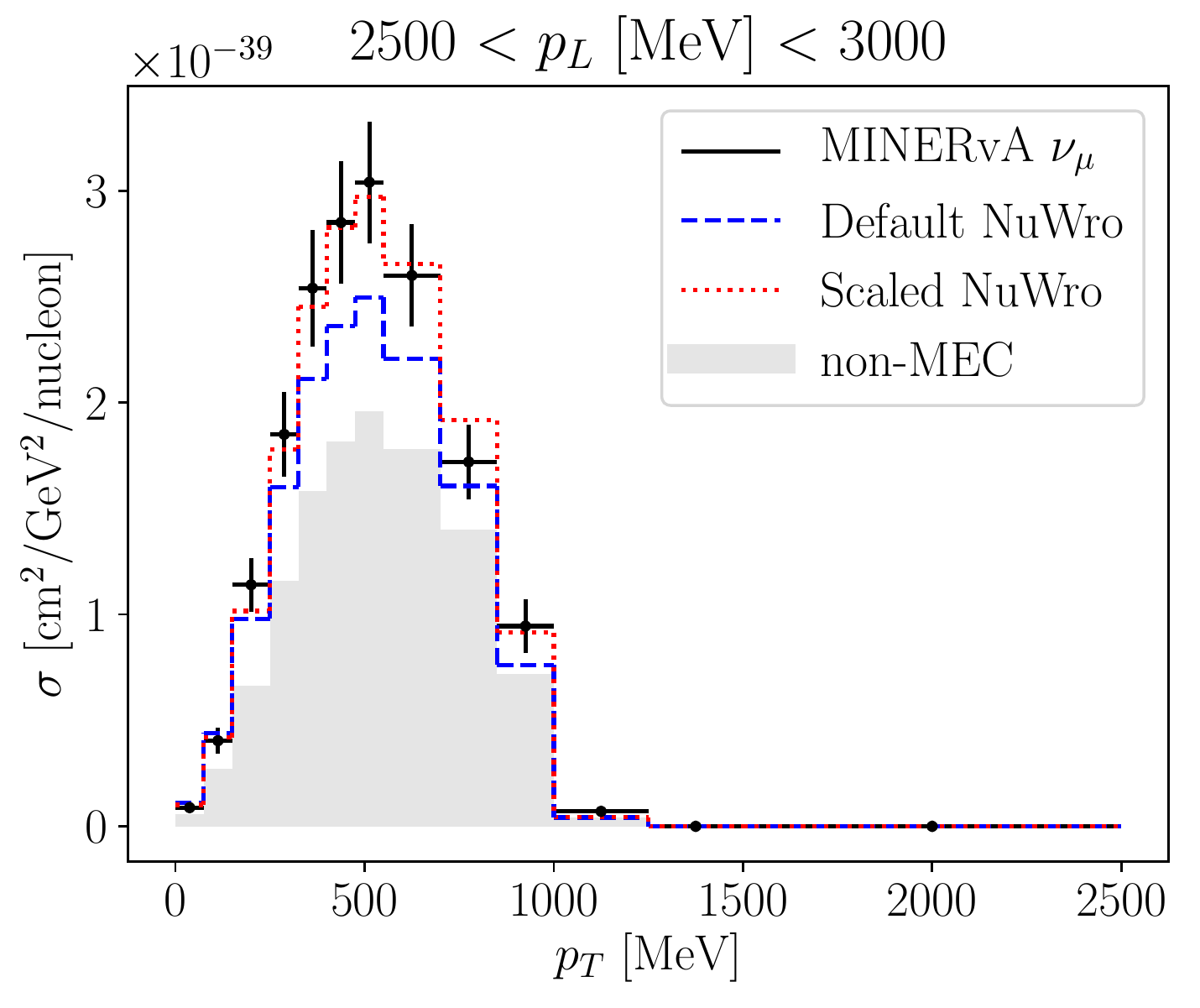}
    \end{subfigure}%
    \medskip
    \begin{subfigure}{.49\textwidth}
      \centering
      \includegraphics[width=0.8\linewidth]{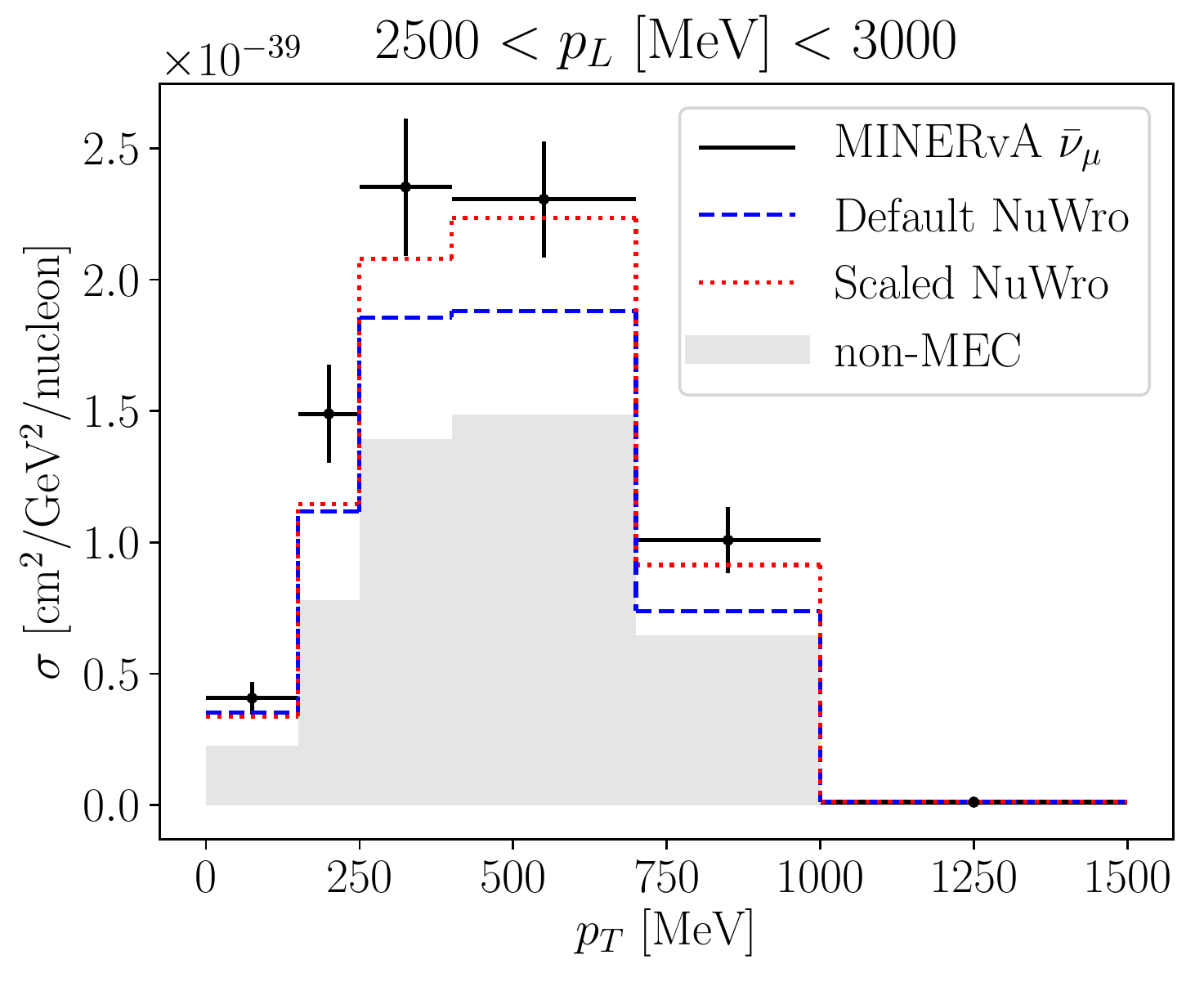}
    \end{subfigure}
    
    \quad
    
    \begin{subfigure}{.49\textwidth}
      \centering
      \includegraphics[width=0.8\linewidth]{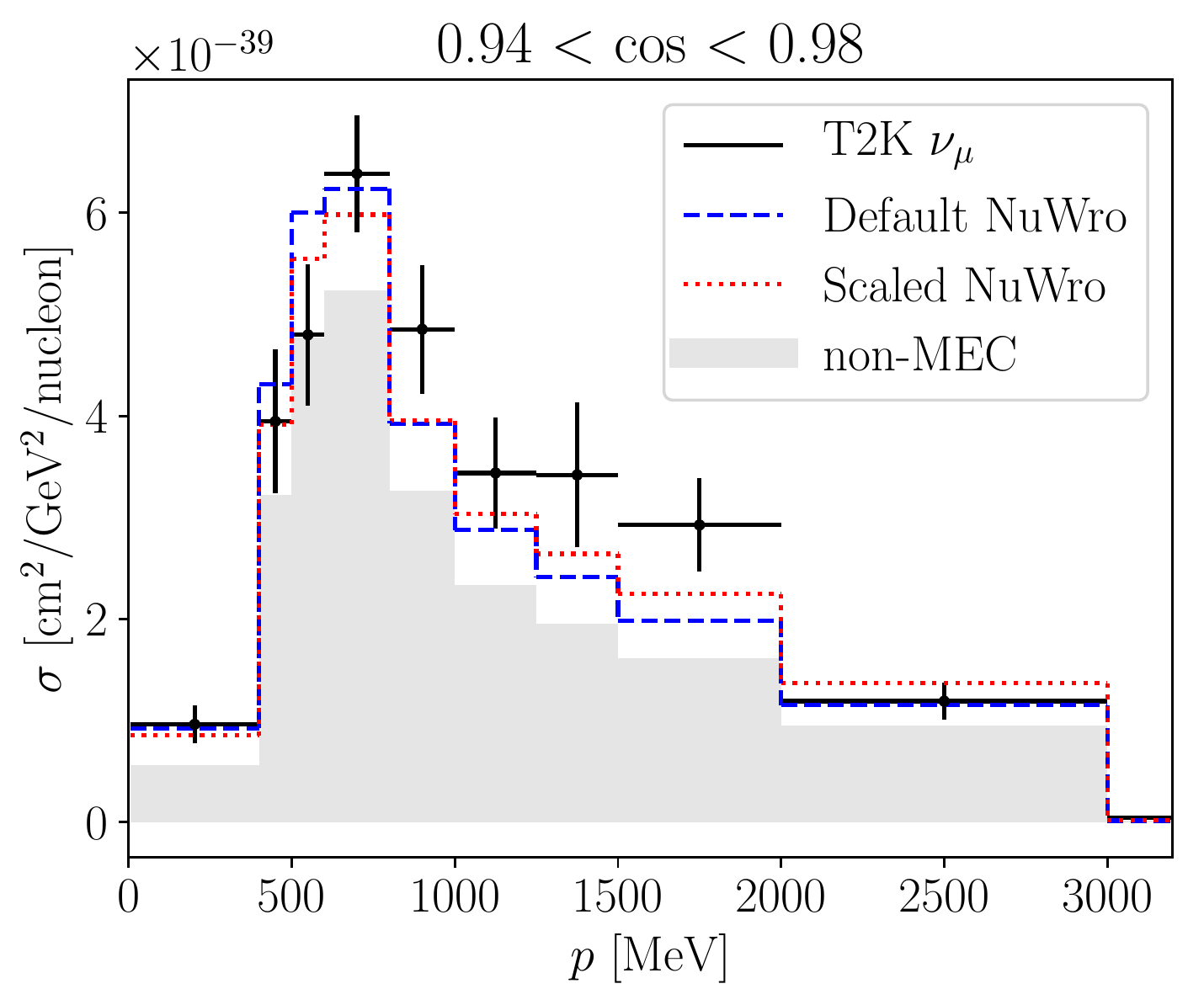}
    \end{subfigure}%
    \medskip
    \begin{subfigure}{.49\textwidth}
      \centering
      \includegraphics[width=0.8\linewidth]{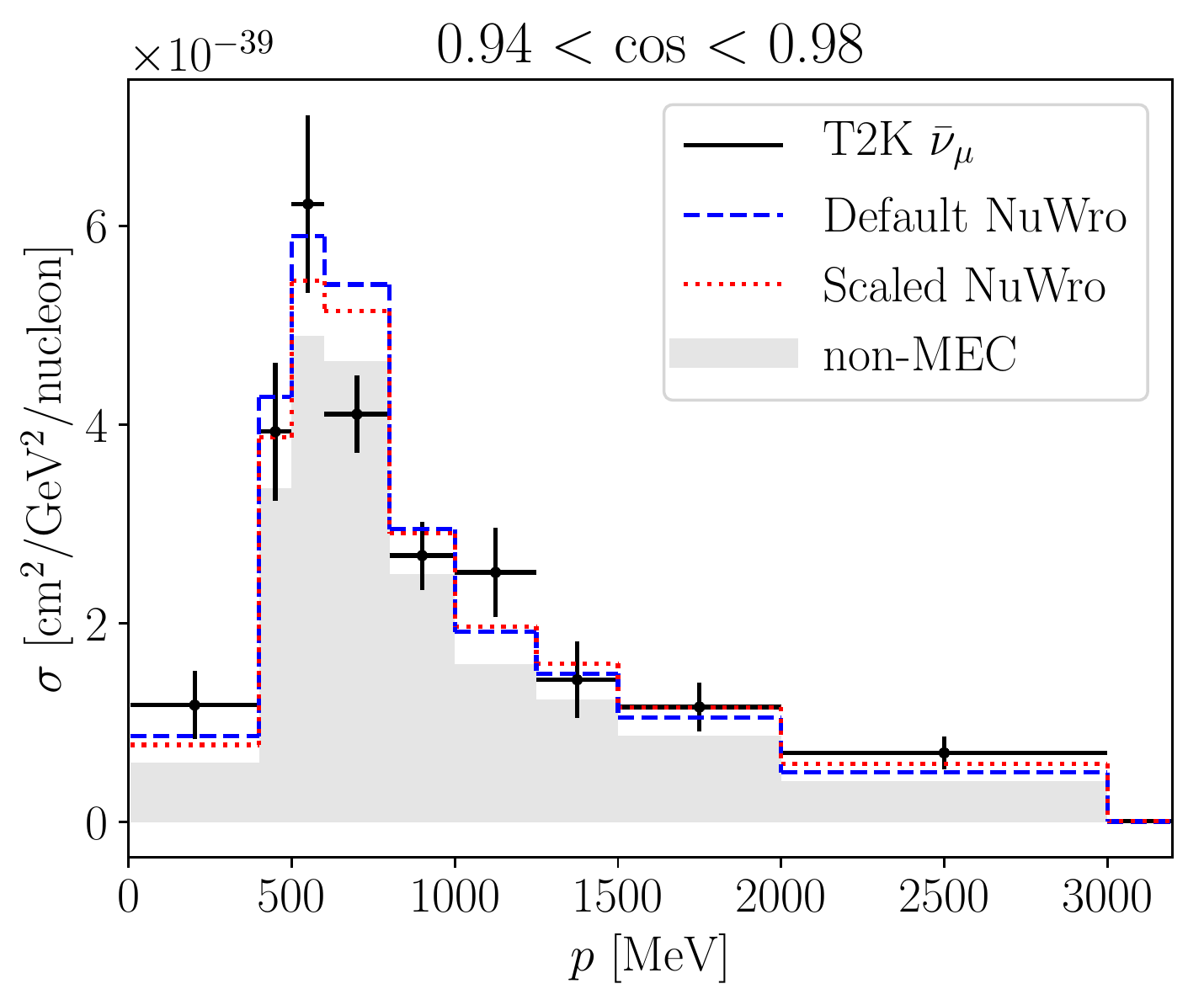}
    \end{subfigure}
    
    \caption{A sample of detail comparison of NuWro predictions before and after rescaling confronted with the experimental data. The shaded area shows a sum of contributions from CCQE, RES and DIS mechanisms.}
    \label{fig:bin_cross_sections}
\end{figure*}

The results shown in Fig \ref{fig:bin_cross_sections} indicate that a significant redistribution of 2p2h cross section is predicted and this translates into a change of values of reconstructed neutrino energy in experiments like T2K where in the Superkamiokande detector final state nucleons are not observed. 

It is interesting that at larger neutrino energies the overall 2p2h cross sections strongly exceed those of  the original Valencia model and seem to be close to the predictions from Martini et al model \cite{Marteau:1999jp, Martini:2009uj} and also SUSAv2 model \cite{Gonzalez-Jimenez:2014eqa}. We obtained also a strong increase of the values of response functions at the boundary of the Valencia model domain i.e. close to $q=1.2$~GeV/c. This may be a signal that the definition of the boundary proposed in Ref.~\cite{Gran:2013kda} is too restrictive and should be relaxed as it is in the SUSAv2 model. In a very recent paper of the Valencia group \cite{Sobczyk:2020dkn} it is argued that there is a large 3p-3h contribution neglected in the original papers. This makes the overall np-nh Valencia model cross section larger and closer to our final result.

The results presented in this paper are the first step in our program of construction of the phenomenological model of 2p2h. The final goal is very involved numerically and we decided to divide it into steps. The final  step is to rescale three most important response matrices in an independent way. $W_3$ enters the cross section formula in Eq.~\ref{eq:inclusive} with different signs for neutrinos and antineutrinos and the results obtained in this paper suggest that it should be scaled in a different way than others. $W_2$ is multiplied by a neutrino energy dependent function that takes different values in MINERvA and T2K experiments. If we allow $W_1$ and $W_2$ to be scaled independently we get extra flexibility to adjust better to both data sets. Altogether, we think that with three independent rescalings we will obtain more reduction of $\chi^2_{final}$ from all the individual experiments. 

In this paper we used the default NuWro version with LFG nucleus model. However, we think that the most prospective will be to use one of the models which are known to reproduce well the QE peak resulting from one-body mechanism, see Ref.~\cite{Sobczyk:2017mts}. One of them, the hole spectral function (SF) approach \cite{Benhar:1994hw}\cj{}{,} is already implemented in NuWro. With CCQE events modeled with SF and three independent rescalings of $W_{1,2,3}$ using the approach described in this paper we should obtain a realistic model of 2p2h contribution on the carbon target. This work is in progress.

\newcommand{\mywidth}{0.75}

\begin{figure*}[h!]
    \vspace{-0.5cm}
    \centering
    \begin{subfigure}{.49\textwidth}
      \centering
      \includegraphics[width=\mywidth\linewidth]{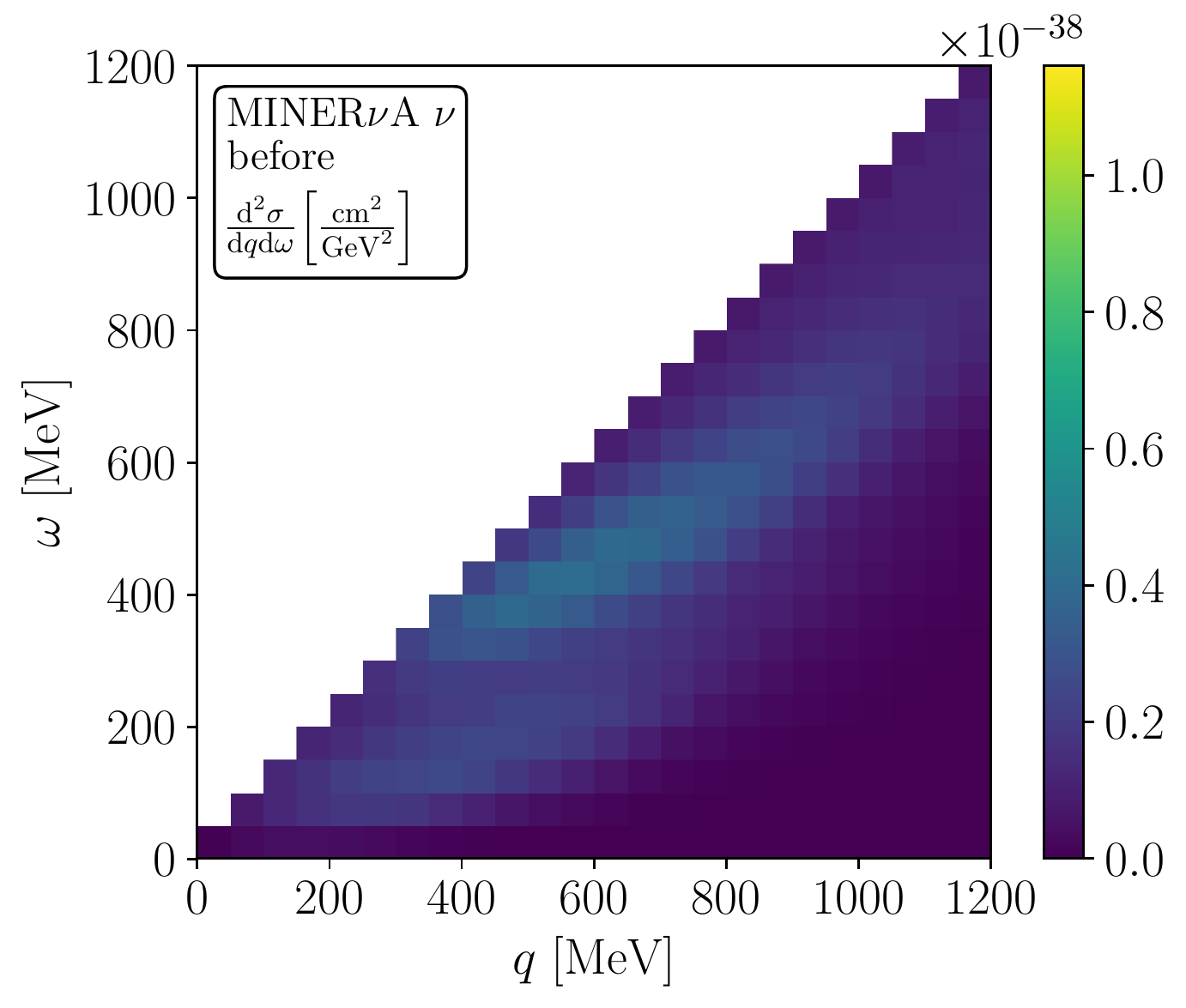}
    \end{subfigure}%
    \begin{subfigure}{.49\textwidth}
      \centering
      \includegraphics[width=\mywidth\linewidth]{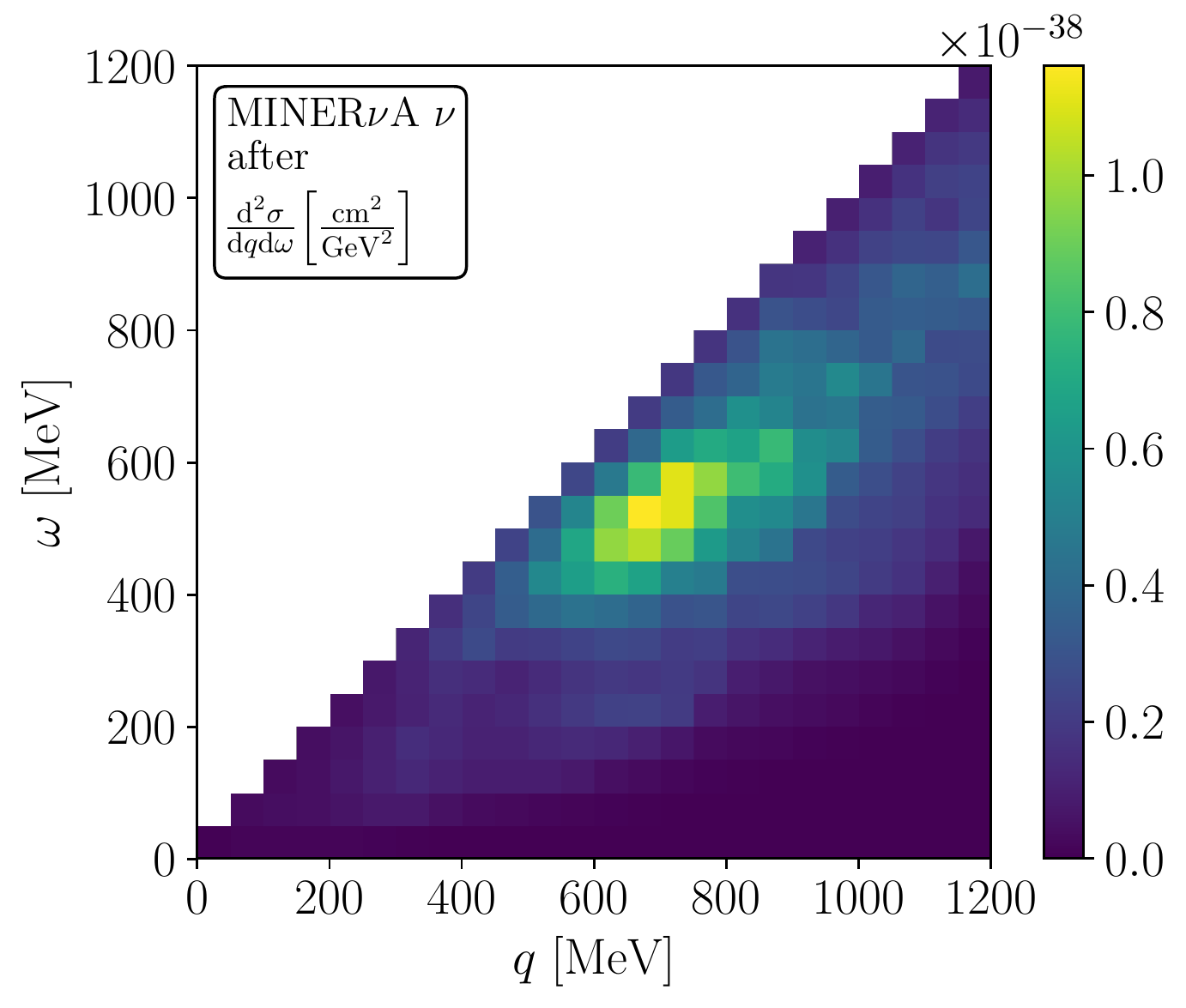}
    \end{subfigure}
    \quad
    \begin{subfigure}{.49\textwidth}
      \centering
      \includegraphics[width=\mywidth\linewidth]{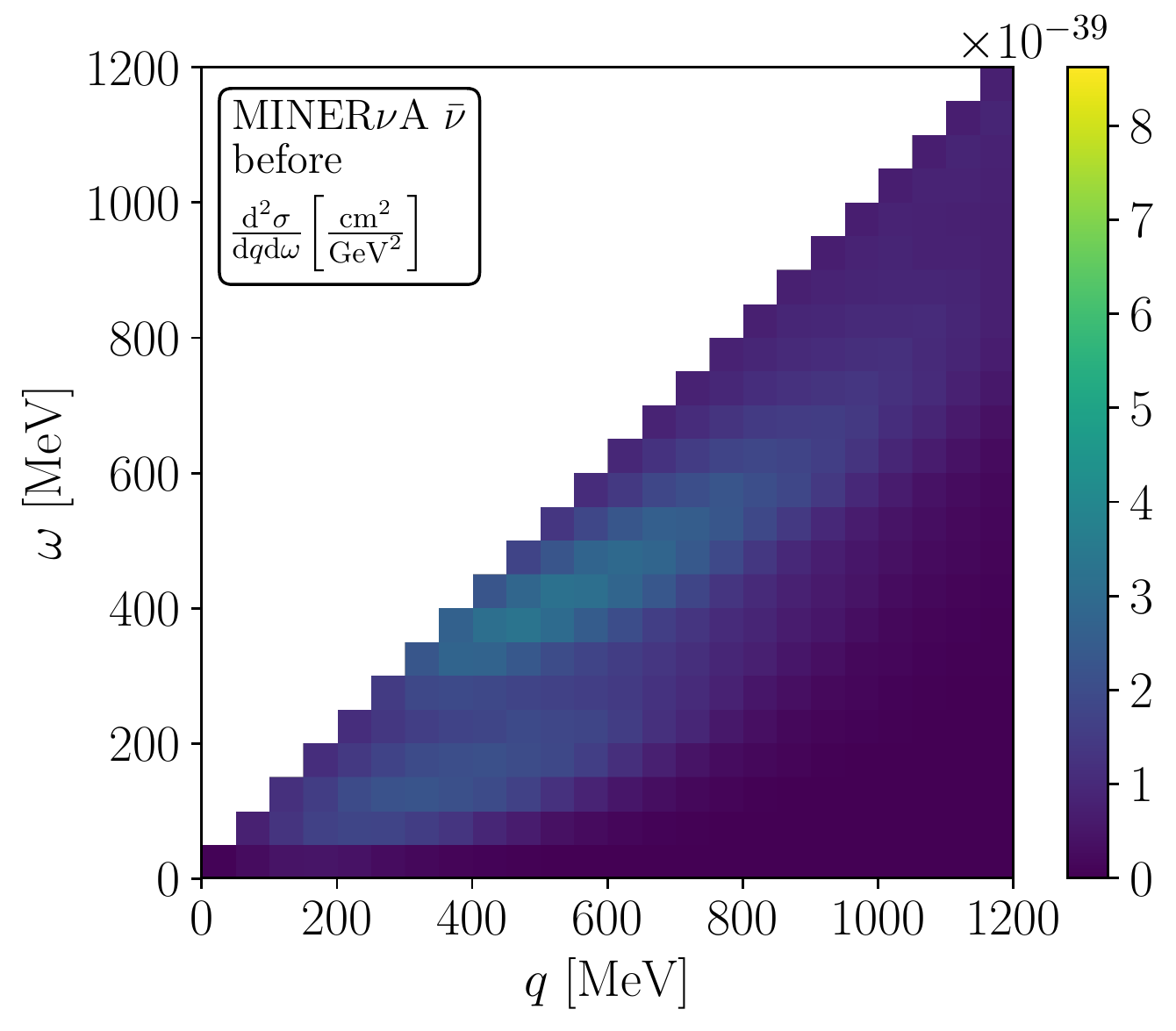}
    \end{subfigure}%
    \begin{subfigure}{.49\textwidth}
      \centering
      \includegraphics[width=\mywidth\linewidth]{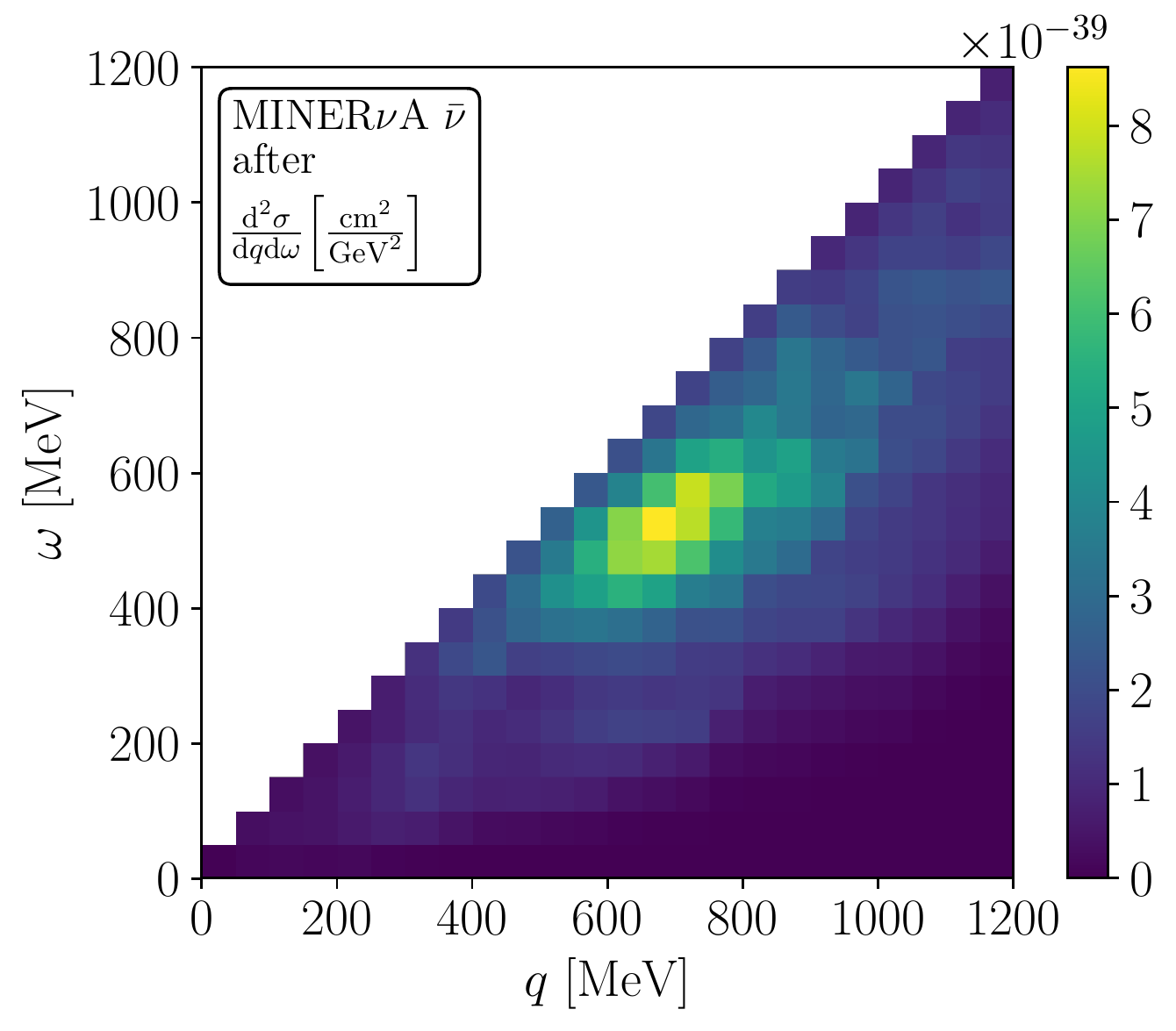}
    \end{subfigure}
    \quad
    \begin{subfigure}{.49\textwidth}
      \centering
      \includegraphics[width=\mywidth\linewidth]{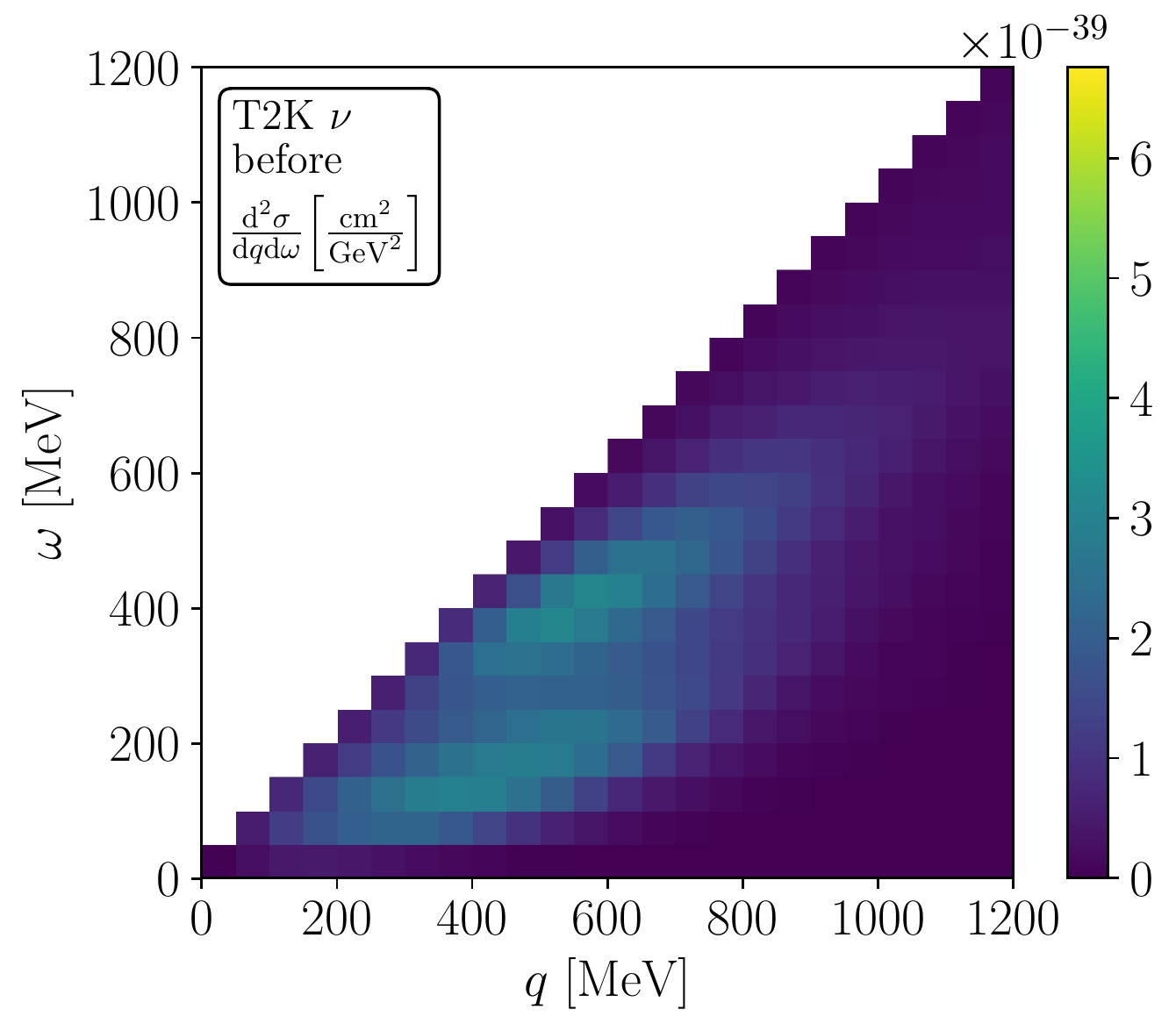}
    \end{subfigure}%
    \begin{subfigure}{.49\textwidth}
      \centering
      \includegraphics[width=\mywidth\linewidth]{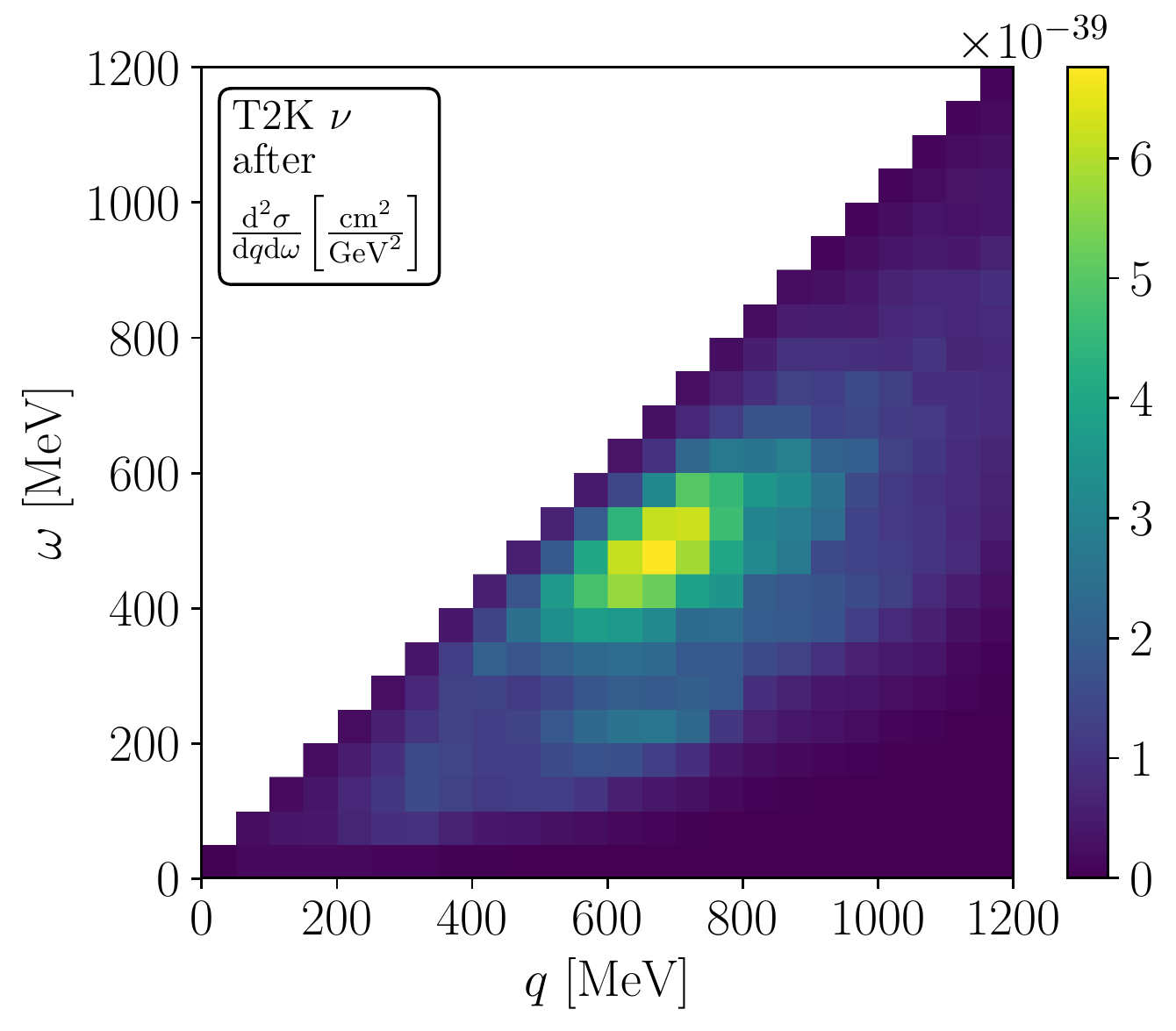}
    \end{subfigure}
    \quad
    \begin{subfigure}{.49\textwidth}
      \centering
      \includegraphics[width=\mywidth\linewidth]{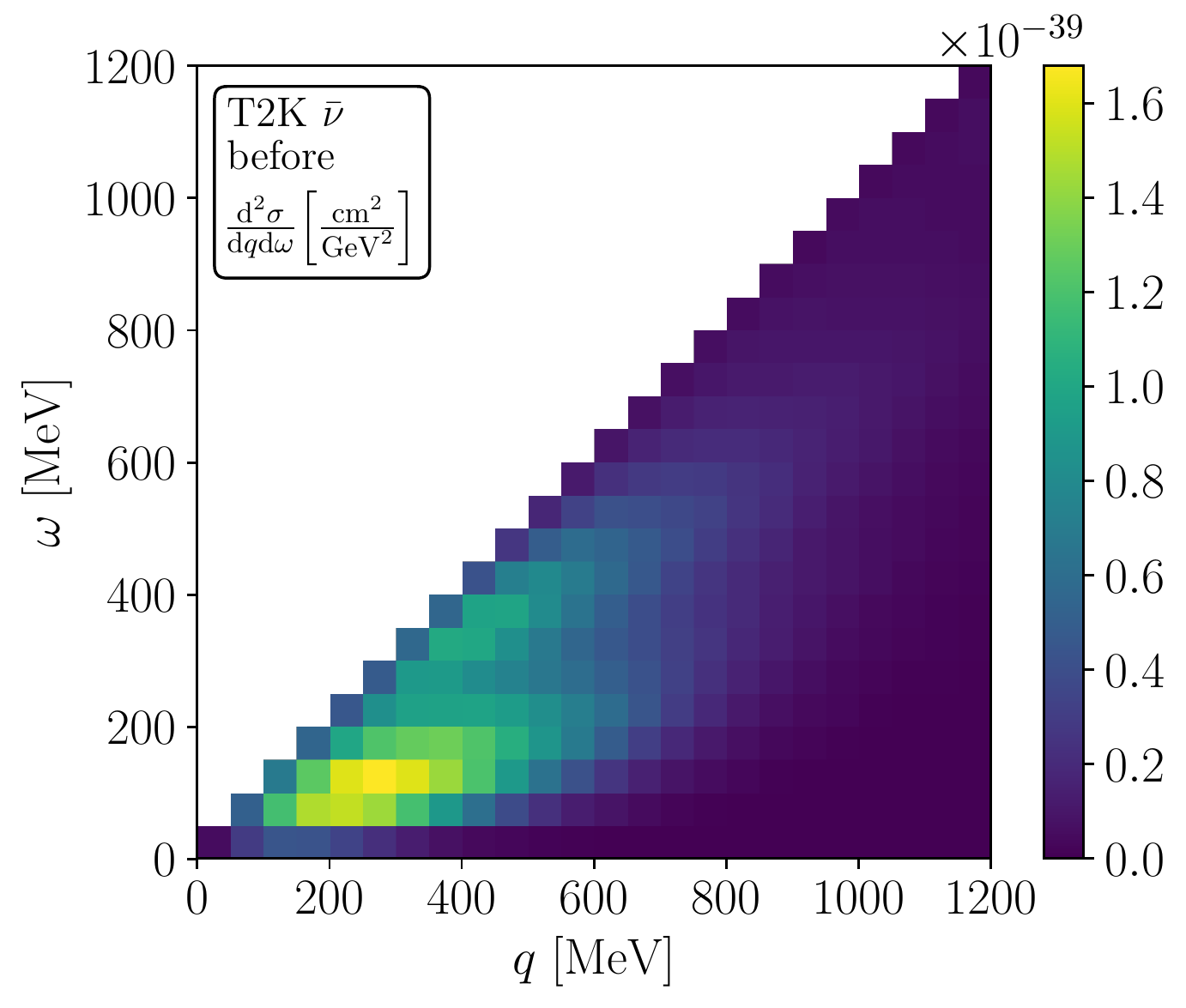}
    \end{subfigure}%
    \begin{subfigure}{.49\textwidth}
      \centering
      \includegraphics[width=\mywidth\linewidth]{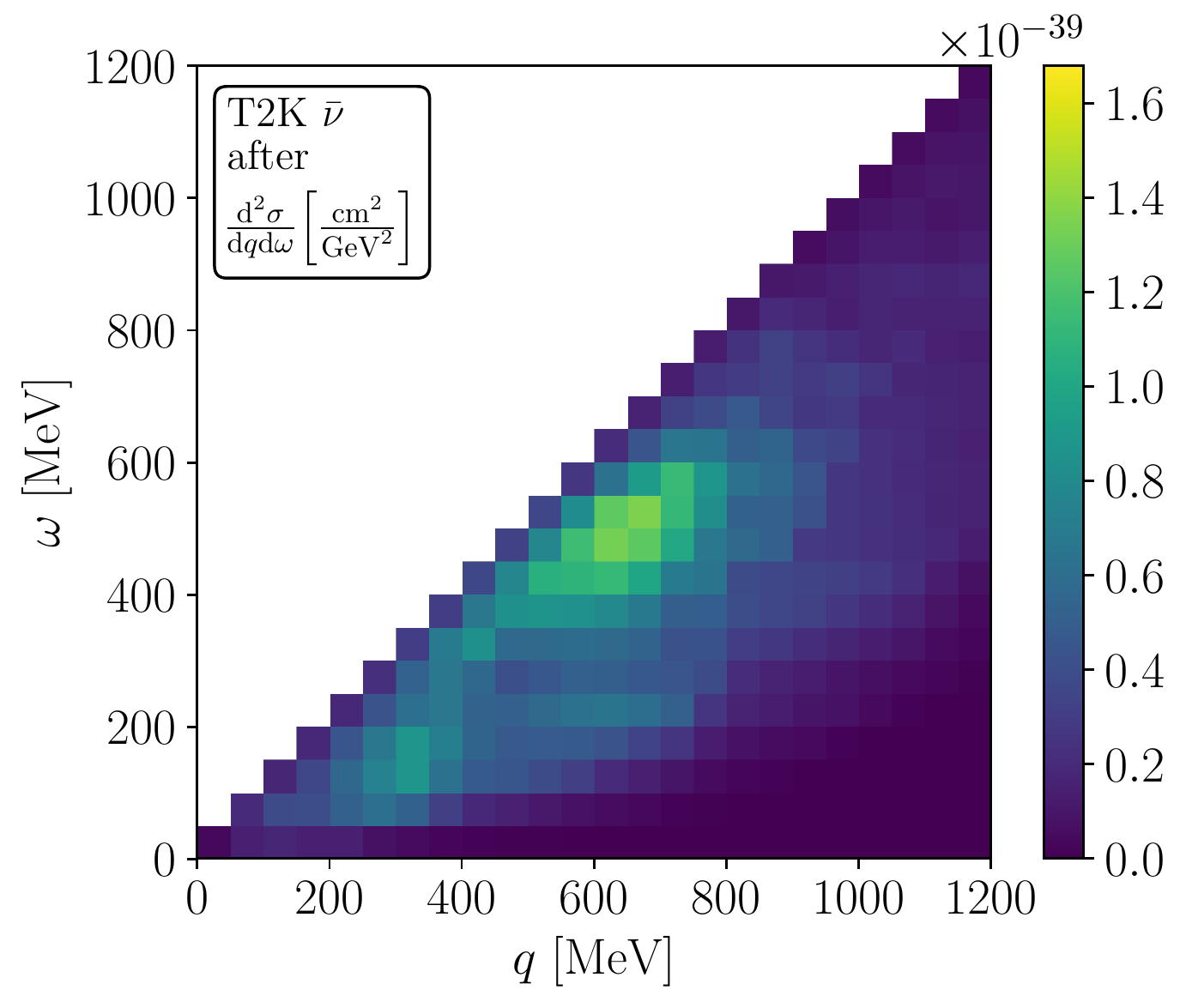}
    \end{subfigure}
    \caption{Distribution of 2p2h cross section in the $\omega/q$ plane before (left) and after (right) rescaling. From top to bottom: MINER$\nu$A neutrino, MINER$\nu$A antineutrino, T2K neutrino and T2K antineutrino.}
    \label{fig:wqplane}
\end{figure*}


\appendix
\section{A decomposition of covariance matrix}

In \cite{Aguilar-Arevalo:2013dva} one can find a procedure how to decompose an arbitrary $N \times N$ covariance matrix into a sum of  `shape`, `normalization` and `mixed` parts:

\begin{widetext}
\begin{eqnarray} V_{jk}=V_{jk}^{shape}+ V_{jk}^{mixed}+V_{jk}^{norm},
\end{eqnarray}

\begin{eqnarray}
V_{jk}^{shape}&=&V_{jk}-\frac{x_k}{x_T}\sum_{l=1}^N V_{jl} - \frac{x_j}{x_T}\sum_{l=1}^N V_{lk}+\frac{x_j x_k}{x_T^2}\sum_{l,s=1}^N V_{ls},\nonumber
\\
V_{jk}^{mixed}&=&\frac{x_k}{x_T}\sum_{l=1}^N V_{jl} + \frac{x_j}{x_T}\sum_{l=1}^N V_{lk}-2\frac{x_jx_k}{x_T^2}\sum_{l,s=1}^N V_{ls},\nonumber
\\
V_{jk}^{norm}&=&\frac{x_jx_k}{x_T^2}\sum_{l,s=1}^N V_{ls},
\end{eqnarray}
\end{widetext}
where $x_j$ are results of measurements and 

\begin{equation}
x_T=\sum_{l=1}^N x_l.\nonumber
\end{equation}

The matrix $V_{jk}^{shape}$ is singular: a vector made of N identical numbers is an eigenvector to eigenvalue 0. This makes the use of $V_{jk}^{shape}$ in the definition of the modified $\chi^2$ difficult. We propose to 
introduce:

\begin{eqnarray} \chi^2_{shape}\equiv \sum_{j,k=1}^N (\tilde y_j-x_j) V^{pseudo}_{shape; j,k} (\tilde y_k-x_k).
\label{eq:pseudochi2}
\end{eqnarray}
where $V^{pseudo}_{shape}$ is a Moore-Penrose pseudoinverse \cite{Penrose:1955vy} of $V^{shape}$. It is a generalization of the definition of inverse matrix, inverse and pseudoinverse matrices coincide for nonsingular matrices. $\tilde y_j$ are normalized to satisfy 

\begin{eqnarray}
\sum_j\tilde y_j=\sum_j x_j.
\end{eqnarray}

We investigated the  statistical properties of the estimator defined in Eq.~\ref{eq:pseudochi2}. We used the covariance matrix of the MINERvA neutrino experiment studied in this paper. We produced several throws $(y_1, ..., y_N)$ generated with a multivariate distribution defined by $(x_1, ..., x_N)$ and $V$. For each one we calculated a `normalized random throws`  $(\tilde y_1, ..., \tilde y_N)$
obtained by applying a normalization factor  $f=\frac{\sum_j x_j}{\sum_j y_j}$: $\tilde y_j=f\cdot y_j$. Finally, we studied a  distribution of values of $\chi^2_{shape}$. It has the basic features of the standard $\chi^2(N-1)$ distribution. The difference is that the peak is less pronounced with more probability at both smaller and larger values of the random variable. It may be difficult to infer from $\chi^2_{shape}$ confidence intervals but it can be used safely as an estimator in a search for best fit values.

\section{A toy model}

In this appendix the performance of $\chi^2$ introduced in Sec.~\ref{sec:estimator} is tested with a simple toy model. Numerical values are chosen to be similar to those used in the Ref.~\cite{D'Agostini:1993uj}.

Suppose two measurements were done with the following results: $
  x=\left[ \begin{array}{c}
   8.0 \\
   8.5 \\
  \end{array}  \right]
  $
and the covariance matrix is reported to be:
\begin{eqnarray}
V=
  \left[ \begin{array}{cc}
   0.6656 & 0.68 \\
   0.68 & 0.7514 \\
  \end{array}  \right]
\end{eqnarray}
Suppose also that a theoretical model predictions contains a parameter $\lambda$ the value of which we would like to estimate based on the data. The model predictions for the two measurements are assumed to be:
  
\begin{eqnarray}
y(\lambda)=\left[ \begin{array}{c}
   7.2+\lambda\cdot 0.795 \\
  7.2+\lambda\cdot 0.805  \\
  \end{array}  \right].
\end{eqnarray}
The standard $\chi^2(\lambda)$ estimator is defined as

\begin{eqnarray}
\chi^2(\lambda)\equiv \sum_{j,k=1}^2 (y_j(\lambda)-x_j)V^{-1}_{jk}(y_k(\lambda)-x_k)
\end{eqnarray}
It can be checked that $\chi^2(\lambda)$ has a minimum at \hbox{$\tilde\lambda = 0.94$} and $
  y(\tilde\lambda)=\left[ \begin{array}{c}
   7.95 \\
   7.96 \\
  \end{array}  \right].$ 
When we compare those values with the measurements we see that we obtained a puzzling result, a manifestation of the
Peelle's  Pertinent  Puzzle \cite{PPP}.

Applying the procedure outlined in the Appendix A we obtain:
\begin{eqnarray}
V^{shape}\approx 0.01359\cdot 
  \left[ \begin{array}{cc}
   1 & -1 \\
   -1 & 1 \\
  \end{array}  \right].
\end{eqnarray}
A pseudoinverse of $V^{shape}$ is:
\begin{eqnarray}
V^{pseudo}_{shape}\approx 18.40\cdot 
  \left[ \begin{array}{cc}
   1 & -1 \\
   -1 & 1 \\
  \end{array}  \right].
\end{eqnarray}

In the $\chi^2_{shape}$ introduced in the Appendix A there is no information about the overall normalization of data points. A remedy is to add a ${\cal N}$ term defined as 

\begin{eqnarray}
{\cal N}=(\sum_{j=1}^2 y_j(\lambda)-\sum_{j=1}^2x_j)^2/\sigma^2_{norm}
\end{eqnarray}

with
\begin{eqnarray}
\sigma^2_{norm}=\sum_{j,k=1}^2 V_{j,k}\approx 2.777.
\end{eqnarray}


Finally we define:

\[ 
\chi^2_{final}(\lambda)\equiv \sum_{j,k=1}^2 (y_j(\lambda)-x_j)\tilde V^{-1,pseudo}_{jk}(y_k(\lambda)-x_k)+ {\cal N}.
\]

It can be checked that $\tilde\chi^2_{final}(\lambda)$ has a minimum at $\tilde\lambda\approx 1.396$ and 
$
  y(\tilde\lambda)=\left[ \begin{array}{c}
   8.31 \\
   8.32 \\
  \end{array}  \right]
  $ which is a reasonable result.


\begin{acknowledgments}
We thank Kajetan Niewczas, Steven Dolan, Sara Bolognesi, Ciro Riccio, Kevin McFarland and other members of the T2K Neutrino Interactions Working Group for many helpful comments and stimulating discussions. 

The authors were supported by the Polish Ministry of Science and Higher Education, Grant DIR/WK/2017/05 and also by NCN Opus Grant 2016/21/B/ST2/01092.
\end{acknowledgments}

\bibliography{refs}

\end{document}